\documentclass[showpacs,10pt,twocolumn,prb]{revtex4-1}

\usepackage{amsmath}
\usepackage{amssymb}
\usepackage{graphicx}
\usepackage{amssymb}
\usepackage{graphics}
\usepackage{epsfig}
\usepackage{CJK}
\usepackage{color}

\begin{document}


\title{Enhanced Thermoelectric Power and Electronic Correlations in RuSe$_2$}
\author{Kefeng Wang,$^{1,\ast,\S}$ Aifeng Wang,$^{1}$ A. Tomic,$^{1,2}$ Limin Wang,$^{1}$  A. M. Milinda Abeykoon,$^{3}$ E. Dooryhee,$^{3}$ S. J. L. Billinge$^{1,2}$ and C. Petrovic$^{1,\ddag}$}
\affiliation{$^{1}$Condensed Matter Physics and Materials Science Department, Brookhaven
National Laboratory, Upton, NY 11973\\
$^{2}$Department of Applied Physics and Applied Mathematics, Columbia University, New York, NY 10027 \\
$^{3}$Photon Sciences Directorate, Brookhaven National Laboratory, Upton, NY 11973}
\date{\today}

\begin{abstract}
We report the electronic structure, electric and thermal transport properties of Ru$_{1-x}$Ir$_{x}$Se$_2$ ($x \leq 0.2$). RuSe$_2$ is a semiconductor that crystallizes in a cubic pyrite unit cell. The Seebeck coefficient of RuSe$_2$ exceeds -200 $\mu$V/K around 730 K. Ir substitution results in the suppression of the resistivity and the Seebeck coefficient, suggesting the removal of the peaks in density of states near the Fermi level. Ru$_{0.8}$Ir$_{0.2}$Se$_{2}$ shows a semiconductor-metal crossover at about 30 K. The magnetic field restores the semiconducting behavior. Our results indicate the importance of the electronic correlations in enhanced thermoelectricity of RuSb$_{2}$.
\end{abstract}

\maketitle

Recent interest in thermoelectric energy conversion induces a wide interest in the materials with  high thermoelectric performance.\cite{TE1,TE2} A narrow distribution or a large peak in the electronic density of states (DOS) close to the Fermi level has long been considered favorable for a high Seebeck coefficient ($S$).\cite{TE3,TE4} Such peak could be induced by the resonant level dopants in semiconductors\cite{TE5,TE6} or by the magnetic interaction between the local magnetic moment and itinerant electrons in many \textit{f}- and \textit{d}-electron based materials\cite{correlation1,magnetic1} It has been reported that some strongly correlated metals (such as heavy fermion metals) and correlated semiconductors (such as Kondo insulators) show significant enhanced Seebeck coefficient and power factor. For example, large peaks in Seebeck coefficient up to 800 $\mu V/K$ were observed in heavy fermion metal CePd$_3$,\cite{CePd3-1,CePd3-2}  Kondo insulator FeSi,\cite{FeSi-1,FeSi-2,FeSi-3} Ce$_3$Pt$_3$Sb$_4$,\cite{CePtSb} and CeFe$_4$P$_{12}$.\cite{CeFeP}

More recently, a very large Seebeck coefficient ($\sim 4\times 10^4 \mu$V/K) and huge power factor ($\sim 2\times 10^3 \mu W/K^2cm$) \cite{FeSb2-1,FeSb2-2,FeSb2-3,FeSb2-4} was observed in FeSb$_2$.\cite{FeSb2-5} This makes the mechanism of thermopower enhancement in FeSb$_2$ of high interest.\cite{FeSb2-6,FeSb2-7,FeSb2-8,FeSb2-9,FeSb2-10,Tomczak}  The results from density functional theory without electron correlation effect can only qualitatively reproduce the temperature dependence of the Seebeck coefficient. The predicted peak value of $S$ is only one tenth of the experimental value, suggesting the importance of strong electronic correlations.\cite{FeSb2-8,FeSi-3,Tomczak}

\begin{figure}[tbp]
\includegraphics[scale=0.3]{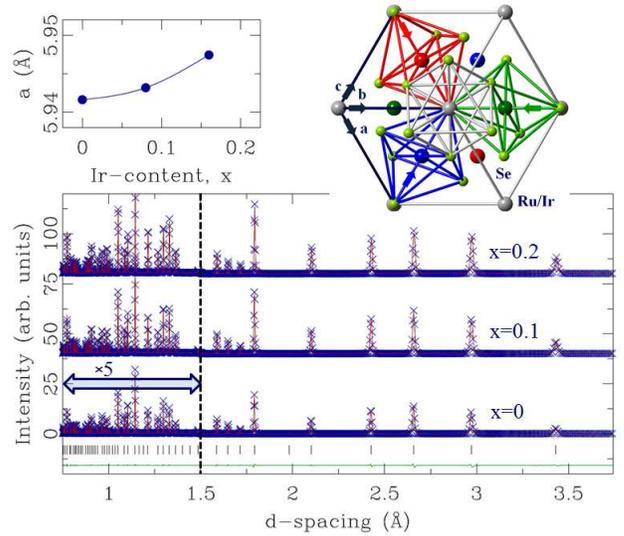}
\caption{(Color online) Ru$_{1-x}$Ir$_{x}$Se$_{2}$ crystal structure considerations.
Main panel shows 300 K experimental x-ray diffraction patterns
(cross symbols) and refined structural model \textit{Pa-3} (solid red line),
with difference curve (green solid line) offset for clarity for x=0.
Vertical ticks mark reflections. Data and model profiles for samples
with x=0.1 and 0.2 nominal composition are included and offset for clarity.
Top left: evolution of lattice parameter with refined Ir content,
solid line is guide for the eyes. Top right: pyrite-type crystal structure
as seen along (111) direction, and distorted RuSe$_{6}$ corner shared octahedra
representing basic building blocks of the structure.}
\end{figure}

Pyrite FeS$_2$ is a semiconductor with a band gap of (0.8 - 0.95) eV and a high light absorbtion.\cite{Seehra,Ennaoui} FeS$_2$ shows a Seebeck coefficient up to $\sim -300 \mu$V/K at 300 K.\cite{Willeke} The Seebeck coefficient of iron dichalcogenides FeX$_2$ [X=(S,Se,Te)] decreases for heavier chalcogens due to decreasing energy gap, but retains relatively large values of $\sim|(2-3)|\cdot10^{2} \mu$V/K above 200 K.\cite{Harada}

Here we report the detailed electronic structure, electric and thermal transport properties of pure and Ir-doped RuSe$_2$ pyrite. RuSe$_2$ shows a semiconducting behavior with an indirect gap from resistivity $\sim 1.5$ eV. The band structure calculation shows a pileup of states near the Fermi level suggesting correlation effects. The Seebeck coefficient of RuSe$_2$ exceeds -200 $\mu$V/K at 730 K, showing the electron-type carriers. Ir doping introduces lattice expansion and extra electrons, which results in the significant suppression of the resistivity and the Seebeck coefficient. The sample with $20\%$ Ir doping shows a semiconductor-metal transition at about 20 K, while the magnetic field restores the semiconducting behavior.

(Ru$_{1-x}$Ir$_x$)Se$_2$ (with $x=0, 0.1, 0.2$) polycrystals were made using a high-temperature solid state reaction method. Stoichiometric Ru (99.99$\%$), Ir (99.99$\%$) and Se (99.9999$\%$) were ground, pelletized, sealed in a quartz tube, heated to 1000 $^{\circ}C$, kept for about 20 hours and then the furnace was turned off. Next, the material was ground, pelletized again and heated with the similar temperature profile at 1100 $^{\circ}C$.  Medium resolution room temperature X-ray diffraction measurements were carried out using a (0.25$\cdot$0.25) mm$^{2}$ 48 keV ($\lambda$ = 0.02487 nm) focused (on the detector) X-ray beam at 28-ID-C beam line at National Synchrotron Light Source II at Brookhaven National Laboratory. The X-ray energy was selected using a horizontally focused double crystal Laue monochromator, with vertical focusing achieved using 1 m long Si mirror. Finely pulverized samples were filled into 1 mm diameter cylindrical polyimide capillaries, and the data collection was carried out in transmission geometry using Perkin Elmer amorphous silicon area detector mounted orthogonal to the beam path 1272.6 mm away from the sample. The raw 2D data were integrated and converted to intensity versus scattering angle using the software Fit2D.\cite{Hammersley}  The average structure was assessed through Rietveld refinements\cite{Rietveld} to the raw diffraction data using the General Structure Analysis System (GSAS)\cite{Larson} operated under EXPGUI,\cite{Toby} utilizing Pa-3 model from the literature.\cite{Lutz} Electrical transport measurements  were conducted on polished samples in Quantum Design PPMS-9 with conventional four-wire method. Thermal transport properties were measured in Quantum Design PPMS-9 from 2 K to 350 K, and in Ulvac ZEM-3 system at higher temperatures, both using one-heater-two-thermometer method. The relative error of each measurement was $\frac{\Delta \kappa}{\kappa}\sim$5$\%$ and $\frac{\Delta S}{S}\sim$10$\%$ based on standard, however at 350 K the discrepancy in measured values was $25\%$. As opposed to ULVAC ZEM-3, PPMS $S$ and $\rho$ were obtained in separate measurements using TTO and ACT option on the same sample. Hence, ULVAC ZEM-3 data were normalized to PPMS values at 300 K. Fist principle electronic structure calculation were performed using experimental lattice parameters within the full-potential linearized augmented plane wave (LAPW) method\cite{wien2k1} implemented in WIEN2k package.\cite{wien2k2} The general gradient approximation (GGA) of Perdew \textit{et al}.,\cite{gga} was used for exchange-correlation potential. The LAPW sphere radius were set to 2.5 Bohr for all atoms. The converged basis corresponding to $R_{min}k_{max}=7$ with additional local orbital were used where $R_{min}$ is the minimum LAPW sphere radius and $k_{max}$ is the plane wave cutoff.

Diffraction data for all three compositions are well explained within pyrite-type \textit{Pa-3} structure, comprised of 3D network of distorted (squashed) Se6 octahedra that coordinate Ru/Ir. Irregularity of the octahedra is reflected in principal axes deviating from 90 degrees (see Figure 1). Refined structural parameters are summarized in Table 1. Lattice parameter increases on substituting Ru with larger Ir. Average Ru-Se near neighbor distance decreases slightly with doping, whereas departure for regularity in octahedral angles increases slightly. Debye-Waller factors increase slightly with doping as well, consistent with presence of quenched disorder introduced by chemical substitution. The Ir doping limit is $20\%$ and above that the synthesis resulted in mixed phases of \textit{Pa-3} space group of pure RuSe$_2$ and $Pnma$ space group of pure IrSe$_2$.

\begin{table*}[b]
\caption{Lattice Parameters for Ir-doped RuSe$_2$}
\begin{ruledtabular}
\begin{tabular}{ccccccc}
Nominal Composition & Lattice(${\AA}$) & $x_{Se}$ & Uiso-Ru ($\AA$) & Uiso-Se ($\AA$) & Refined Composition & R($\%$) \\
RuSe$_2$ & 5.94164(3) & 0.3807(2) & 0.0039(3) & 0.0051(3) & Ru$_{0.99(1)}$Se$_2$ & 7.3 \\
Ru$_{0.9}$Ir$_{0.1}$Se$_2$ & 5.94319(7) & 0.3794(2) & 0.0042(3) & 0.0058(3) & Ru$_{0.92(1)}$Ir$_{0.08(1)}$Se$_2$ & 5.6 \\
Ru$_{0.8}$Ir$_{0.2}$Se$_2$ & 5.94740(7) & 0.3784(2) & 0.0045(3) & 0.0061(3) & Ru$_{0.84(1)}$Ir$_{0.16(1)}$Se$_2$ & 4.5 \\
\hline
\hline
Nominal Composition & Se-Ru ($\AA$) & Se-Ru-Se ($^o$) & Se-Ru-Se ($^o$) && & \\
RuSe$_2$ & 2.4742(3) & 94.71(2) & 85.29(2) & & & \\
Ru$_{0.9}$Ir$_{0.1}$Se$_2$ & 2.4722(3) & 94.82(2) & 85.18(2) & & & \\
Ru$_{0.8}$Ir$_{0.2}$Se$_2$ & 2.4720(3) & 94.91(2) & 85.09(2) & & &
\end{tabular}
\end{ruledtabular}
\end{table*}

\begin{figure}[tbp]
\includegraphics[scale=0.6]{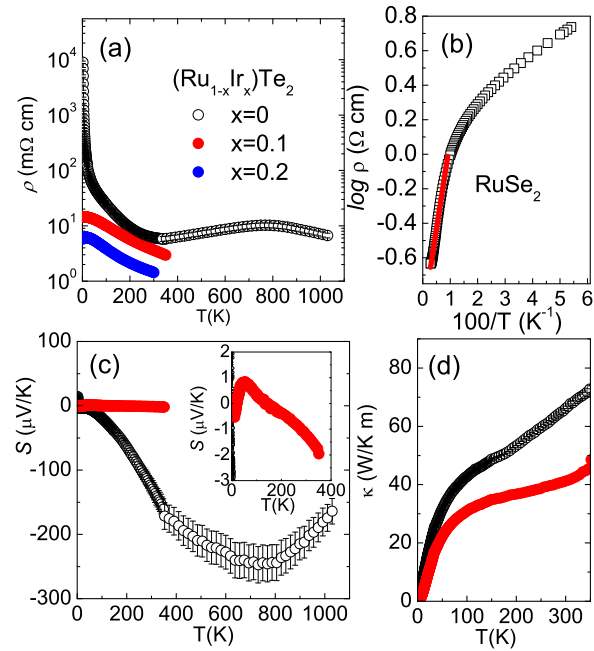}
\caption{(Color online) The Resistivity $\rho$ (a), the $\log \rho$ vs $1/T$ relationship for pure RuSe$_2$ and the solid line is the linear fitting result using thermal activation theory (b), Seebeck coefficient $S$ (c) and thermal conductivity $\kappa$ (d) for (Ru$_{1-x}$Ir$_x$)Se$_2$ with $x=0, 0.1$ and 0.2.}
\end{figure}

Fig. 2 shows the electrical and thermal transport properties. The resistivity of RuSe$_2$ [Fig. 2(a)]) shows typical semiconducting behavior. The fitting for thermal activation conductivity [the solid line in Fig. 2(b)] estimates that the main band gap is $\sim 1.5$ eV. The slope of the resistivity or the energy gap changes in the low temperature range, possibly due to the native $d$-states or  impurity states within the main band gap, similar to Fe$_{1-x}$Ru$_{x}$Sb$_{2}$.\cite{Fuccillo} The Seebeck coefficient [Fig. 2(c)] approaches 180 $\mu$V/K at $350$ K and shows a peak of about 247 $\mu$V/K at about 730 K. The thermal conductivity of RuSe$_2$ [Fig. 2(d)] is rather high in the whole range of measured temperatures.

The semiconducting behavior of RuSe$_{2}$ is consistent with the first principle calculation results (Fig. 3). The density of states [Fig. 3(a)] shows a gap with size of 0.4 eV, whereas the band structure [Fig. 3(b)] indicates that RuSe$_2$ is a indirect-gap semiconductor. However, the gap size from the density functional theory (0.4 eV) is much smaller than the transport gap (about 1.5 eV), suggesting that the electron correlations may be important. There are several narrow bands just below the Fermi level with Ru \textit{4d} orbital character [the heavy lines in Fig 3(b)]. This is confirmed by the large pileup of states [Fig. 3(a)]. But there is also significant hybridization between Ru \textit{4d} and Se \textit{p} orbitals indicated by the overlap between the peaks from Ru and Se below and above the Fermi level.

The high Seebeck coefficient of RuSe$_{2}$  should come from the peaks in density of states just below the Fermi level [Fig. 3(a)], with major contribution from the narrow $d$-bands. Although the Seebeck coefficient of RuSe$_2$ is high, its power factor is small because of the high resistivity.

\begin{figure}[tbp]
\includegraphics[scale=0.8]{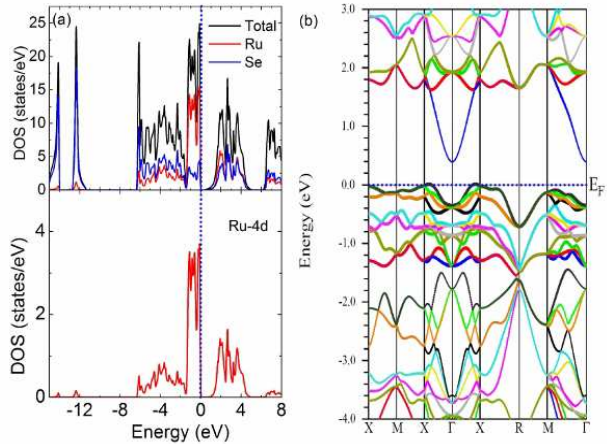}
\caption{(Color online) (a) The density of states and (b) the band structure of RuSe$_2$. The different colors in band structure indicate different bands and the thickness of the bands represents the weight of Ru 4d orbitals.}
\end{figure}

Ir doping introduces carriers and is effective in enhancing the conductivity, however it also significantly suppresses Seebeck coefficient [Fig. 2(a,c)]. The $10\%$ Ir doped sample still shows semiconducting behavior but the residual resistivity decreases by two orders of magnitude at 2 K and in half at 200 K when compared to pure RuSe$_2$. Further increase in Ir substitution suppresses the resistivity even more. The sample with $20\%$ Ir doping shows a semiconductor-metal crossover at $\sim 30$ K; below that temperature the resistivity begins decreasing with decreasing temperature [Fig. 2(a) and Fig. 4(a)]. The thermal conductivity is also suppressed by Ir doping [Fig. 2(d)], possibly due to the lattice disorder introduced by the doping. The $10\%$ Ir doping reduces the Seebeck coefficient to only about 2 $\mu$V/K at 300 K [inset in Fig. 2(c)].

\begin{figure}[tbp]
\includegraphics[scale=0.7]{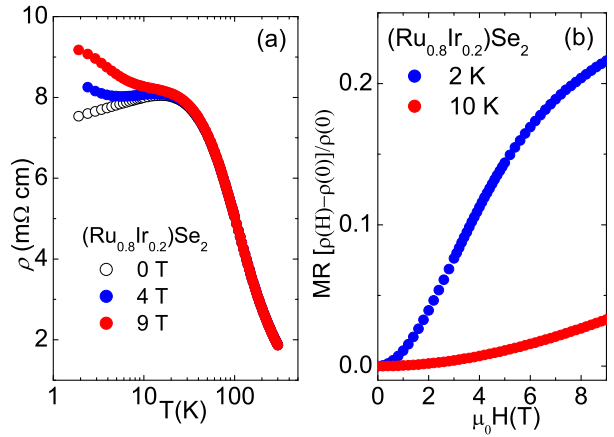}
\caption{(Color online) (a) Temperature dependence of the resistivity $\rho$ of (Ru$_{0.8}$Ir$_{0.2}$)Se$_2$ in different magnetic fields. (b) The magnetoresistace ratio $MR=(\rho(H)-\rho(0))/\rho(0)$ as the function of the magnetic field for Ru$_{0.8}$Ir$_{0.2}$)Se$_2$ with $T=$ 2 K and 10 K, respectively. }
\end{figure}

The magnetic field has significant influence on the transport of Ru$_{0.8}$Ir$_{0.2}$Se$_2$ in the low temperature range  [Fig. 4(a)]. Above the semiconductor-metal crossover at $\sim 30$ K, the application of magnetic field has minute effect on the resistivity. But below it, in the metallic regime, the magnetic field enhances the value of resistivity and changes its temperature dependence. In 4 T field, the $\rho$(T) still undergoes the semiconductor-metal transition at the same temperature but then changes to semiconducting behavior below 4 K. The 9 T magnetic field totally smears out the semiconductor-metal transition and restores the semiconducting behavior. The MR is always positive at 2 K and tends to saturate in higher fields [Fig.4(b)].

The clear suppression of the value and the slope of the resistivity indicates that the Ir-doping induces the decrease of the band gap. The $10\%$ Ir doping only introduces 0.1 electrons per unit cell. This will result in a slight shift of the Fermi level toward higher energy direction [Fig. 3(a)], within the framework of the density functional theory. Since $10\%$ Ir-doped RuSe$_2$ is still a semiconductor, it is reasonable to believe that the Fermi level is still in the gap. If so, this slight shift of the Fermi level could not induce the huge suppression of the Seebeck coefficient [Fig. 2(c)]. Since the Seebeck coefficient is related to the energy slope of the density of states near the Fermi level, this implies the effect of electronic correlations. The $5d$ electrons of Iridum feature less localized (i.e. more extended) wave functions when compared to $4d$ electrons of Ruthenium. Hence, Iridium substitution not only introduces extra carriers, but also reduces the electronic correlations. Furthermore, since the magnetic field restores the semiconducting behavior in sample with $20\%$ Ir level, the original semiconducting behavior in pure RuSe$_2$ could be related to some extent to the magnetic mechanism. Taken together with enhanced thermopower, our results suggest that phyical properties of RuSe$_2$ may share some similarity with the correlated electron semiconductor FeSb$_{2}$.\cite{FeSb2-4,FeSb2-5} The comparison of the electronic structure, electronic and magnetic correlations of pyrite RuSe$_2$ to marcasite FeSb$_{2}$ could be important for studies of structural effects on correlated electron thermoelectricity and deserves further studies.

Electronic structure of pyrites such as NiS$_{2-x}$Se$_{x}$ is related to the occupation of \textit{d} orbitals which have significant influence on the band filling and correlation effect.\cite{Imada} Pyrites and marcasites both feature distorted octahedral coordination of transition metals (e.g. Fe or Ru) in the local structure.\cite{Goodenough} Whereas the octahedra share common corners in the cubic pyrite unit cell and the resulting crystal field at Ru in RuSe$_{2}$ has trigonal symmetry, the orthorombic marcasite unit cell features linear chains of edge-sharing octahedra parallel to orthorombic c-axis. In both cases \textit{d}-electrons dominate the electronic states near the Fermi level. In RuSe$_{2}$ the t$_{2g}$ orbitals are completely filled as opposed to t$_{2g}$ orbitals in marcasite FeSb$_{2}$.\cite{Goodenough,Brostigen} This inhibits a possibility for thermally induced anisotropic metallic states and thermally induced enhanced Pauli susceptibility.\cite{FeSb2-4,FeSb2-5,Petrovic2} On the other hand, when comparing RuSe$_{2}$ to RuSb$_{2}$, Se(\textit{4p4}) has an extra electron when compared to Sb (\textit{5p3}). So it is expected that the occupation of \textit{d} orbitals in RuSe$_{2}$ is different from RuSb$_{2}$ which would change correlation strength. This is reflected in the density of states: Ru d-states in RuSe$_{2}$ (Fig. 3) are more enhanced near the Fermi level when compared to RuSb$_{2}$.\cite{Fuccillo}  Triangular arrangement of metal (i.e. Ru) atoms in pyrite lattice of RuSe$_{2}$ (Fig. 1) could enhance the correlation effects in RuSe$_{2}$ even further via geometric frustration. This has been theoretically considered\cite{Katsura} and experimentally verified in pyrite NiS$_{2}$.\cite{Matsuura} Since geometrical frustration coupled with strong Coulomb interaction may enhance thermoelectric power,\cite{Gu} putative antiferromagnetic states in RuSe$_{2}$ materials are of interest. The ZT=S$^{2}$/$\rho$$\kappa$ value at 300 K (730 K) is only about 0.003 (0.005 assuming about 100 W/Km). Thermal conductivity is rather high and very far away from the amorphous limit. Nanoengineering of RuSe$_{2}$ objects may reduce thermal conductivity and could lead to much larger values of ZT.\cite{Hochbaum,Ren,Kanatzidis,WanC}

In conclusion, we report enhanced thermoelectric power and electronic correlations in RuSe$_2$ RuSe$_2$ shows a semiconducting behavior with a thermally activated gap. The band structure calculation confirmed the semiconducting characteristic, albeit with significantly underestimated gap implying the importance of the electron correlation effect. The Seebeck coefficient of RuSe$_2$ approaches - 250 $\mu V/K$ near 730 K, showing the electron-type carriers. Small Ir doping results in the significant suppression of the resistivity and the Seebeck coefficient. The sample with $20\%$ Ir doping shows a semiconductor-metal transition at about 20 K, while the magnetic field restores the semiconducting behavior at low temperature. Our results shows the large Seebeck coefficient of RuSe$_2$ and implies the important role of electron and magnetic correlations.

\begin{acknowledgments}
Work at Brookhaven is supported by the U.S. DOE under contract No. DE-AC02-98CH10886. X-ray scattering data were collected at 28-ID-C x-ray powder diffraction beam line at National Synchrotron Light Source II at Brookhaven National Laboratory.
\end{acknowledgments}

$^{\ast }$Present address: Department of Physics, University of Maryland, College
Park, MD 20742-4111, USA.
$^{\S}$ wangkf@umd.edu
${\ddag }$ petrovic@bnl.gov


\begin{thebibliography}{99}
\bibitem{TE1}
G. D. Mahan and J. O. Sofo, "The best thermoelectric", Proc. Natl. Acad. Sci. U.S.A. \textbf{93}, 7436-7439 (1996).
\bibitem{TE2}
G. J. Snyder, and E. S. Toberer, "Complex thermoelectric materials", Nature Mater. \textbf{7}, 105-114 (2008).
\bibitem{TE3}
Yanzhong Pei, Xiaoya Shi, Aaron LaLonde, Heng Wang, Lidong Chen and G. Jeffrey Snyder, "Convergence of electronic bands for high performance bulk thermoelectrics." Nature, \textbf{473}, 66-69 (2011).
\bibitem{TE4}
L. E. Bell, "Cooling, heating, generating power and recovering waste heat with thermoelectric system", Science \textbf{321}, 1457-1461 (2008).
\bibitem{TE5}
J. P. Heremans V. Jovovic, E. S. Toberer, A. Sarmat, K. Kurosaki, A. Charoenphakdee, S. Yamanaka and G. Jeffrey Snyder, "Enhancement of thermoelectric efficiency in PbTe by distorion of the electronic density of states", Science \textbf{321}, 554-557 (2008).
\bibitem{TE6}
J. P. Heremans B. Wiendlocha and A. M. Chamoire, "Resonant levels in bulk thermoelectric semiconductors." Energy Environ. Sci. \textbf{5}, 5510-5530 (2012).
\bibitem{correlation1}
G. D. Mahan, Solid State Physics \textbf{51}, 81 (1998).
\bibitem{magnetic1}
J. Kondo, "Giant thermo-electric power of dilute magnetic alloys", Prog. Theor. Phys. \textbf{34}, 372-382 (1965).
\bibitem{CePd3-1}
G. D. Mahan, B. Sales, and J. Sharp, "Thermoelectric Materials: New Approaches to an Old Problem", Phys. Today \textbf{50}, 42 (1997).
\bibitem{CePd3-2}
Y. Ijiri and F. J. DiSalvo, "Thermoelectric properties of R$_x$Ce$_{1-x}$Pd$_3$ (R=Y, La$_{0.5}$Y$_{0.5}$, Nd)", Phys. Rev. B \textbf{55}, 1283 (1998).
\bibitem{FeSi-1}
R. Wolfe, J. H. Wernick, and S. E. Haszko, "Thermoelectric properties of FeSi." Phys. Lett. \textbf{19}, 449-450 (1965).
\bibitem{FeSi-2}
Brian C. Sales, Olivier Delaire, Michael A. McGuire, and Andrew F. May, "Thermoelectric properties of Co-, Ir-, and Os-doped FeSi alloys: Evidence for strong electron-phonon coupling", Phys. Rev B \textbf{83}, 125209 (2011).
\bibitem{FeSi-3}
J. M. Tomczak, K. Haule and G. Kotliar, "Signatures of elecronic correlations in iron silicide", Proc. Natl. Acad. Sci. U.S.A. \textbf{109}, 3243-3246 (2012).
\bibitem{CePtSb}
C. D. W. Jones, K. A. Regan, and F. J. DiSalvo, "Thermoelectric properties of the doped Kondo insulator: Nd$_x$Ce$_{ 3-x}$Pt$_3$Sb$_4$," Phys. Rev. B \textbf{58}, 16057 (1998).
\bibitem{CeFeP}
H. Sato \textit{et al}., "Anomalous transport properties of RFe4P12 (R = La, Ce, Pr, and Nd)" Phys. Rev. B \textbf{62}, 15125 (2000).
\bibitem{FeSb2-1}
A. Bentien, S. Johnsen, G. K. H. Madsen, B. B. Iversen, and F. Steglich, "Colossal Seebeck coefficient in strongly correlated semiconductor FeSb$_2$," Euro. Phys. Lett. \textbf{80}, 17008 (2007).
\bibitem{FeSb2-2}
P. Sun, N. Oeschler, S. Johnsen, B. B. Iversen, and F. Steglich, "FeSb$_2$: Prototype of huge electron-diffusion thermoelectricity," Phys. Rev. B \textbf{79}, 153308 (2009).
\bibitem{FeSb2-3}
P. Sun, N. Oeschler, S. Johnsen, B. B. Iversen, and F. Steglich, "Huge thermoelectric power factor: FeSb$_2$ versus FeAs$_2$ and RuSb$_2$," Appl. Phys. Express \textbf{2}, 091102 (2009).
\bibitem{FeSb2-4}
Qing Jie, Rongwei Hu, Emil Bozin, A. Llobet, I. Zaliznyak, C. Petrovic, and Q. Li, "Electronic thermoelectric power factor and metal-insulator transition in FeSb$_{2}$", Phys. Rev. B \textbf{86}, 115121 (2012).
\bibitem{FeSb2-5}
C. Petrovic Y. Lee, T. Vogt, N. Dj. Lazarov, S. L. Bud'ko and P. C. Canfield, "Kondo insulator description of spin state transition in FeSb$_2$," Phys. Rev. B \textbf{72}, 045103 (2005).
\bibitem{FeSb2-6}
Huaizhou Zhao M. Pokharel, Gaohua Zhu, Shuo Chen, K. Lukas, Qing Jie, C. Opeil, Gang Chen and Zhifeng Ren "Dramatic thermal conductivity reduction by nanostructures for large increase in thermoelectric figure-of-merit of FeSb$_2$," Appl. Phys. Lett. \textbf{99}, 163101 (2011).
\bibitem{FeSb2-7}
Kefeng Wang, Rongwei Hu, J. Warren, and C. Petrovic, "Enhancement of the thermoelectric properties in doped FeSb$_2$ bulk crystals," J. Appl. Phys. \textbf{112}, 013703 (2012).
\bibitem{FeSb2-8}
M. Koirala Huaizhou Zhao, M. Pokharel, Shuo Chen, T. Dahal, C. Opeil, Gang Chen and Zhifeng Ren, "Thermoelectric property enhancement by Cu nanoparticles in nanostructured FeSb$_2$," Appl. Phys. Lett. \textbf{102}, 213111 (2013).
\bibitem{FeSb2-9}
A. Bentien, G. K. H. Madsen, S. Johnsen, and B. B. Iversen, "Experimental and theoretical investigations of strongly correlated FeSb$_{2-x}$Sn$_x$," Phys. Rev. B \textbf{74}, 205105 (2006).
\bibitem{FeSb2-10}
P. Sun, N. Oeschler, S. Johansen, B. B. Iversen and F. Steglich, "Narrow band gap and enhanced thermoelectricity in FeSb$_2$," Dalton Trans. \textbf{39}, 1012-1019 (2010).
\bibitem{Tomczak} J. Tomczak K. Haule, A. Georges and G. Kotliar, "Thermopower of correlated semiconductors: Application to FeAs$_{2}$ and FeSb$_{2}$", Phys. Rev. B \textbf{82}, 085104 (2010).
\bibitem{Seehra} M. S. Seehra and S. S. Seehra, "Temperature dependence of the band gap of FeS$_{2}$ Phys. Rev. B \textbf{12}, 6620 (1979).
\bibitem{Ennaoui} A. Ennaoui S. Fiechter, Ch. Pettenkofer, N. Alonso-Vante, K. Buker, M. Bronold, Ch. Hopfner and H. Tributsch, "Iron disulfide for solar energy conversion", Sol. Energ. Mat. Sol. Cells \textbf{29}, 085104 (1993).
\bibitem{Willeke} G. Willeke, O. Blenk, Ch. Kloc and E. Bucher, "Preparation and electrical transport of pyrite (FeS$_{2}$ single crystals", J. Alloys Comp. \textbf{178}, 181 (1992).
\bibitem{Harada} T. Harada, "Transport properties of Iron Dichalcogenides FeX$_{2}$ (X=S,Se and Te)", J. Phys. Soc. Jpn. \textbf{67}, 1352 (1998).
\bibitem{Hammersley} A. P. Hammersley, S. O. Svenson, M. Hanfland, and D. Hauserman, "Two-dimensional detector sotware: From real detector to idealised image of two-theta scan", High Pressure Res. \textbf{14}, 235 (1996).
\bibitem{Rietveld} H. M. Rietveld, "Line profiles of neutron powder-diffraction peaks for structure refinement", Acta Crystallogr. \textbf{22}, 151 (1967).
\bibitem{Larson} A. C. Larson and R. B. Von Dreele, "General structure analysis system," (1987), report No. LAUR-86-748, Los Alamos National Laboratory, Los Alamos, NM 87545.
\bibitem{Toby} B. H. Toby, "EXPGUI, a graphical user interface for GSAS", J. Appl. Crystallogr. \textbf{34}, 201 (2001).
\bibitem{Lutz} H. D. Lutz, B. Muller, T. Schmidt, and T. Stingl, "Structure Refinement of Pyrite-Type Ruthenium Disulfide, RuS$_{2}$, and Ruthenium Diselenide RuSe$_{2}$", Acta Crystallogr. C \textbf{46}, 2003 (1990).
\bibitem{wien2k1} M. Weinert, E. Wimmer and A. J. Freeman, "Total-energy all-electron density functional method for bulk solids and surfaces", Phys. Rev. B. \textbf{26}, 4571 (1982).
\bibitem{wien2k2} P. Blaha, K. Schwarz, G. K. H. Madsen, D. Kvasnicka and J. Luitz, "WIEN2k, An Augmented Plane Wave 1 Local Orbitals Program for Calculating Crystal Properties" (Karlheinz Schwarz, Techn. Universitat Wien, Austria), 2001. ISBN
3-9501031-1-2.
\bibitem{gga} J. P. Perdew, K. Burke and M. Ernzerhof, "Generalized Gradient Approximation Made Simple", Phys. Rev. Lett. \textbf{77}, 3865 (1996).
\bibitem{Fuccillo} M. K. Fuccillo, Q. D. Gibson, Mazhar N. Ali, L. M. Schoop and R. J. Cava, "Correlated evolution of colossal thermoelectric effect and Kondo insulating behavior", Appl. Phys. Lett. Materials \textbf{1}, 062102 (2013).
\bibitem{Imada} M. Imada, A. Fujimori and Y. Tokura, "Metal-Insulator Transitions", Rev. Mod. Phys. \textbf{70}, 1039 (1998).
\bibitem{Goodenough} J. B. Goodenough, "Energy Bands in TX$_{2}$ Compounds with Pyrite, Marcasite and Arsenopyrite Structures", J. Solid State Chem. \textbf{5}, 144 (1972).
\bibitem{Brostigen} G. Brostigen and A. Kjeksus, "Bonding Schemes for Compounds with the Pyrite, Marcasite and Arsenopyrite Type Structures", J. Acta Chemica Scandinavica \textbf{24}, 2993 (1970).
\bibitem{Petrovic2} C. Petrovic, J. W. Kim, S. L. Bud'ko, A. I. Goldman, P. C. Canfield, W. Choe and G. J. Miller, "Anisotropy and large magnetoresistance in the narrow-gap semiconductor FeSb$_{2}$", Phys. Rev. B \textbf{67}, 155205 (2003).
\bibitem{Katsura} S. Katsura and T. Imaizumi, "Annealed Ising Bond-Mixture on the Pyrite Lattices", Progress of Theoretical Physics \textbf{67}, 434 (1982).
\bibitem{Matsuura} M. Matsuura, Y. Endoh, H. Hiraka, K. Yamada, A. S. Mishchenko, N. Nagaosa and I. V. Solovyev, "Classical and quantum spin dynamics in the fcc antiferromagnet NiS$_{2}$ with frustration", Phys. Rev. B \textbf{68}, 094409 (2003).
\bibitem{Gu} Xian-Lu Gu, Feng Lu, Da-Yong Liu and Liang-Jian Zou, "Thermoelectric power of single-orbital and two-orbital Hubbard models on triangular lattices", Physica B \textbf{405}, 4145 (2010).
\bibitem{Arsenault} L.-F. Arsenault, B. S. Shastry, P. Semon and A.-M. S. Tremblay, "Entropy, frustration and large thermopower of doped Mott insulators on the fcc lattice", Phys. Rev. B \textbf{87}, 035126 (2013).
\bibitem{Hochbaum} A. I. Hochbaum, Renkun Chen, R.D. Delgado, Wenjie Liang, E. C. Garnett, M. Najarian, A. Majumdar and Peidong Yang, "Enhanced thermoelectric performance of rough silicon nanowires", Nature \textbf{451}, 163 (2007).
\bibitem{Ren} B. Poudel, Q. Hao, Y. Ma, Y. Lan, A. Minnich, B. Yu, X. Yan, D. Wang, A. Muto, D. Vashaee, X. Chen, J. Liu, M. Dresselhaus, G. Chen, and Z. F. Ren, "High Thermoelectric Performance of Nanostructured Bismuth Antimony Telluride Bulk Alloys," Science \textbf{320}, 634 (2008).
\bibitem{Kanatzidis} M. G. Kanatzidis, "Nanostructured Thermoelectrics: The New Paradigm", Chem. Mater. \textbf{22}, 648 (2010).
\bibitem{WanC} L.-F. Chunlei Wan, Yifeng Wang, Ning Wang, Wataru Norimatsu, Michiko Kusunoki and Kunihito Koumoto, "Development of novel thermoelectric materials by reduction of lattice thermal conductivity", Sci. Technol. Adv. Mater. \textbf{11}, 044306 (2010).



\end{thebibliography}
\end{document}